\documentclass{iopart}
\usepackage{euscript,iopams,amssymb,amsfonts,graphicx,bm}
\usepackage{pgfplots,setstack}

\usepackage{float}
\usepackage{epsfig}
\usepackage{esint}

\usepackage{bm,braket}

\eqnobysec

\newcommand{\calU}{{\mathcal U}}
\newcommand{\calG}{{\mathcal G}}
\newcommand{\calP}{{\mathcal P}}
\newcommand{\calQ}{{\mathcal Q}}

\newcommand{\R}{{\mathbb R}}

\newcommand{\X}{\mathbf{X}}
\renewcommand{\P}{\mathbb{P}}
\newcommand{\PP}{\widetilde{P}}
\newcommand{\QQ}{\widetilde{Q}}
\newcommand{\x}{\mathbf{x}}

\renewcommand{\e}{{\mathrm e}}
\newcommand{\E}{{\mathbb E}}

\newcommand{\n}{\mathbf n}
\newcommand{\calT}{{\mathcal T}}
\renewcommand{\P}{\mathbb P}

\newcommand{\ellh}{\hat{\ell}}

\begin{document}

 \title[Diffusion-mediated surface reactions]{Diffusion-mediated surface reactions, Brownian functionals and the Feynman-Kac formula}

\author{Paul C. Bressloff}
\address{Department of Mathematics, University of Utah 155 South 1400 East, Salt Lake City, UT 84112}

\begin{abstract} 
Many processes in cell biology involve diffusion in a domain $\Omega$ that contains a target $\calU$ whose boundary $\partial \calU$ is a chemically reactive surface. Such a target could represent a single reactive molecule, an intracellular compartment or a whole cell. Recently, a probabilistic framework for studying diffusion-mediated surface reactions has been developed that considers the joint probability density or propagator for the particle position and the so-called boundary local time. The latter characterizes the amount of time that a Brownian particle spends in the neighborhood of a point on a totally reflecting boundary. The effects of surface reactions are then incorporated via an appropriate stopping condition for the boundary local time. In this paper we generalize the theory of diffusion-mediated surface reactions to cases where the whole interior target domain $\calU$ acts as a partial absorber rather than the target boundary $\partial \calU$. Now the particle can freely enter and exit $\calU$, and is only able to react (be absorbed) within $\calU$. The appropriate Brownian functional is then the occupation time (accumulated time that the particle spends within $\calU$)
rather than the boundary local time. We show that both cases can be considered within a unified framework by using a Feynman-Kac formula to derive a boundary value problem (BVP) for the propagator of the corresponding Brownian functional, and introducing an associated stopping condition. We illustrate the theory by calculating the mean first passage time (MFPT) for a spherical target $\calU$ located at the center of a spherical domain $\Omega$. This is achieved by solving the propagator BVP directly, rather than using spectral methods. We find that if the first moment of the stopping time density is infinite, then the MFPT is also infinite, that is, the spherical target is not sufficiently absorbing.

\end{abstract}
%%%%%%%%%%%%%%%%%%%%%%%%%%%

\maketitle
\section{Introduction}

\begin{figure}[t!]
  \raggedleft
  \includegraphics[width=12cm]{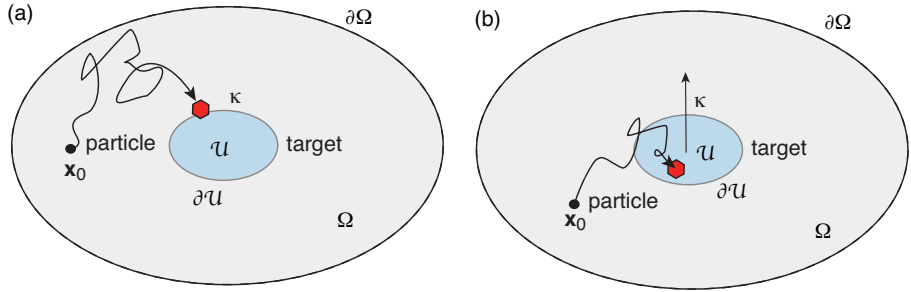}
  \caption{Two models of a reactive target $\calU$ in the interior of a domain $\Omega$. (a) A diffusing particle reacts at a rate $\kappa$ when it is a neighborhood of the target boundary $\partial \calU$. The time spent at the boundary is characterized by the boundary local time. (b) A particle diffuses in and out of the target domain $\calU$ and reacts at a finite rate $\kappa$ within $\calU$; such a reaction could represent transfer to compartment offset from $\Omega$. The time spent within $\calU$ is determined by the occupation time.}
  \label{fig1}
\end{figure}

 Many processes in cell biology involve diffusion in a domain $\Omega$ that contains a target $\calU$ (or possibly multiple targets) whose boundary $\partial \calU$ is a chemically reactive surface, see Fig. \ref{fig1}(a). Such a target could represent a single reactive molecule, an intracellular compartment or a whole cell. One quantity of interest is the Smoluchowski rate at which diffusing particles in the bulk react with the given interior target, which is determined by the net flux into the boundary $\partial \calU$ \cite{Smoluchowski17,Collins49,Rice85,Redner01}.
 If a reaction occurs immediately when a particle first encounters the surface (diffusion-limited), then the surface is totally absorbing and the corresponding boundary condition is Dirichlet. That is, the particle concentration vanishes on the boundary, $c(\x,t)=0$ for $\x \in \partial \calU$. On the other hand, for finite reaction rates there is a nonzero probability that a particle is reflected at the surface and returns to the bulk before a reaction occurs. The surface is then partially absorbing and the typical boundary condition is Robin, that is, 
 $-D\nabla c(\x,t)\cdot \n=\kappa_0 c(\x,t)$, $ \x \in \partial \calU$, 
 where $\n$ is the unit normal at the boundary directed towards the interior of the target, $D$ is the diffusivity, and $\kappa_0$ (in units m/s) is known as the reactivity constant. The totally absorbing case is recovered in the limit $\kappa_0 \rightarrow \infty$, whereas the case of an inert (perfectly reflecting) target is obtained by setting $\kappa_0=0$. In practice, the diffusion-limited and reaction-limited cases correspond to the regimes $\xi \ll R$ and $\xi \gg R$, respectively. Here $R$ is a geometric length-scale that characterizes the size of the target domain $\calU$ and $\xi=D/\kappa_0$ is known as the reaction length. 
 
At the single-particle level, the diffusion equation (or more general Fokker-Planck equation) represents the evolution of the probability density $p(\x,t|\x_0)$ for the particle's position given that it started at $\x_0$. In an unbounded domain, the probability density determines the distribution of random trajectories of a Brownian particle. However, the inclusion of boundary conditions within the probabilistic framework is more complicated. The simplest example is a Dirichlet boundary condition, which can be incorporated into Brownian motion by introducing the notion of a first passage time (FPT), whereby the stochastic process is stopped on the first encounter between the particle and boundary. On the other hand, a totally or partially reflecting boundary requires a modification of the stochastic process itself. For example, a Neumann boundary condition can be implemented in terms of so-called reflected Brownian motion, which involves the introduction of a Brownian functional known as the boundary local time \cite{Levy39,McKean75,Majumdar05}. The latter characterizes the amount of time that a Brownian particle spends in the neighborhood of a point on the boundary. Heuristically speaking, the differential of the local time generates an impulsive kick whenever the particle encounters the boundary, leading to the so-called stochastic Skorokhod equation \cite{Freidlin85}. One can also extend the theory to develop a probabilistic implementation of the Robin boundary condition for partially reflected Brownian motion \cite{Papanicolaou90,Milshtein95} and more general continuous stochastic processes \cite{Singer08}.

Recently, Grebenkov \cite{Grebenkov19b,Grebenkov20,Grebenkov21} has used the boundary local time to develop a theoretical framework for investigating more general forms of diffusion-mediated absorption by partially reactive surfaces. The basic idea is to consider the joint probability density or propagator $P(\x,\ell,t|\x_0)$ for the pair $(\X_t,\ell_t)$ in the case of a perfectly reflecting boundary, where $\X_t$ and $\ell_t$ denote the particle position and local time, respectively. The effects of surface reactions are then incorporated via an appropriate stopping condition for the boundary local time. In particular, the single-particle probabilistic version of the Robin boundary condition (partially reflected Brownian motion) can be implemented 
 by introducing the stopping time 
$
{\mathcal T}=\inf\{t>0:\ \ell_t >\widehat{\ell}\}$,
 with $\widehat{\ell}$ an exponentially distributed random variable that represents a stopping local time \cite{Grebenkov06,Grebenkov07,Grebenkov20}. That is, $\P[\widehat{\ell}>\ell]=\e^{-\gamma\ell}$
with $\gamma=\xi^{-1}=\kappa_0/D$. Since the Robin boundary condition maps to an exponential law for the stopping local time $\widehat{\ell}_t$, the probability density $p(\x,t|\x_0)$ can be expressed in terms of the Laplace transform of the (full) propagator $P(\x,\ell,t|\x_0)$ with respect to the local time $\ell$. 
The advantage of this formulation is that one can consider a more general probability distribution $\Psi(\ell) = \P[\ellh>\ell]$ for the stopping local time $\ellh$ such that \cite{Grebenkov19b,Grebenkov20,Grebenkov21}
$  p(\x,t|\x_0)=\int_0^{\infty} \Psi(\ell)P(\x,\ell,t|\x_0)d\ell$.
This accommodates a wider class of surface reactions where, for example, the reactivity $\kappa(\ell)$ depends on the local time $\ell$ (or the number of surface encounters). Since one can no longer impose a Robin boundary condition for $p(\x,t|\x_0)$, it is necessary to calculate the propagator $P(\x,\ell,t|\x_0)$. This is carried out in Ref. \cite{Grebenkov20} using a non-standard integral representation of the probability density $p(\x,t|\x_0)$ and spectral properties of the so-called Dirichlet-to-Neumann operator.

In this paper we generalize the theory of diffusion-mediated surface reactions to cases where the whole interior target domain $\calU$ acts as a partial absorber rather than the target boundary $\partial \calU$, see Fig. \ref{fig1}(b). Now the particle can freely enter and exit $\calU$, and is absorbed at a rate $\kappa_0$ when inside $\calU$ in the case of constant reactivity. One important example is the passive or active intracellular transport of a vesicle (particle) along the axon or dendrite of a neuron, with absorption corresponding to the transfer of the vesicle to a synaptic target within the surface membrane of the neuron \cite{Bressloff20,Bressloff21,Schumm21}. We show that the main difference between absorption by the target boundary and target interior is that the latter involves the occupation time (accumulated time that the particle spends within $\calU$)
rather than the local time. Otherwise, the theory proceeds along analogous lines to the local time with an associated propagator and a stopping occupation time. In order to calculate the propagator for more general Brownian functionals such as the occupation time, we exploit the fact that the moment generator of a Brownian functional 
satisfies a Feynman-Kac formula \cite{Kac49,Majumdar05}. This allows us to derive a boundary value problem (BVP) for the corresponding propagator $P$ of a general Brownian functional. Taking the latter to be the boundary local time or occupation time of a target and introducing an associated stopping condition then generates the probability density for a Brownian particle undergoing diffusion-mediated surface reactions.

The structure of the paper is as follows. In Sect. 2 we briefly review the theory of surface-mediated reactions developed in Ref. \cite{Grebenkov20} and indicate the natural generalization to absorption within the target domain. The derivation of the BVP for the propagator of a Brownian functional is presented in Sect. 3 and applied to the particular cases of boundary local times and occupation times. The effects of surface reactions are then incorporated via an appropriate stopping condition. In each case, a general formula for the MFPT to be absorbed by the target is constructed in terms of the underlying propagator. In section 4 we explicitly calculate the MFPT for a spherical target $\calU$ located at the center of a spherical domain $\Omega$. That is, $\Omega\backslash \calU$ is a spherical shell whose outer surface is reflecting and whose inner surface is partially reflecting. We exploit the spherical symmetry of the configuration to solve the propagator BVPs in terms of modified Bessel functions. We find that if the stopping time density $\psi=-\Psi'$ has an infinite first moment, then the MFPT is also infinite, that is, the spherical surface is not sufficiently absorbing. Finally, in section 5 we indicate how to extend the analysis to multiple targets.
  
  \section{Diffusion-mediated surface reactions}

Consider a particle diffusing in a bounded domain $\Omega$ containing an interior target $\calU$ with a partially reflecting boundary $\partial \calU$ (Robin boundary condition), see Fig. \ref{fig1}(a). The probability density
$p(\x,t|\x_0)$ satisfies the BVP
\numparts 
\begin{eqnarray}
\label{diffloc1}
\fl 	&\frac{\partial p(\x,t|\x_0)}{\partial t} = D\nabla^2 p(\x,t|\x_0) \mbox{ for } \x\in \Omega\backslash \calU,\quad \nabla p(\x,t|\x_0) \cdot \n=0 \mbox{ for } \x\in\partial \Omega,\\
\fl &D\nabla p(\x,t|\x_0) \cdot \n=-\kappa_0 p(\x,t|\x_0) \mbox{ for } \x\in \partial \calU,\ p(\x,0|\x_0)=\delta(\x-\x_0).
\label{diffloc2}
	\end{eqnarray}
	\endnumparts 
	Here $D$ is the diffusivity and $\kappa_0$ is a constant reactivity. The vector $\n$ represents the unit normal at a boundary point that is directed outwards from the domain $\Omega\backslash \calU$. 
For simplicity, the exterior boundary $\partial \Omega$ is taken to be totally reflecting. Following \cite{Grebenkov19b,Grebenkov20,Grebenkov21}, one can develop a single-particle probabilistic version of the Robin boundary condition in terms of the boundary local time $\ell_t$.	
 Let $\X_t \in \Omega$ represent the position of the particle at time $t$. The boundary local time for a totally reflecting surface $\partial \calU$ is then defined according to \cite{Grebenkov19a}
\begin{equation}
\label{loc}
\ell_t=\lim_{h\rightarrow 0} \frac{D}{h} \int_0^t\Theta(h-\mbox{dist}(\X_{\tau},\partial \calU))d\tau,
\end{equation}
where $\Theta$ is the Heaviside function. Note that $\ell_t$ has units of length due to the additional factor of $D$.
Let $P(\x,\ell,t|\x_0)$ denote the joint probability density or propagator for the pair $(\X_t,\ell_t)$ and introduce the stopping time \cite{Grebenkov19b,Grebenkov20,Grebenkov21} \begin{equation}
\label{Tell}
{\mathcal T}=\inf\{t>0:\ \ell_t >\widehat{\ell}\},
\end{equation}
 with $\widehat{\ell}$ an exponentially distributed random variable that represents a stopping local time \cite{Grebenkov06,Grebenkov07,Grebenkov20}. That is, $\P[\widehat{\ell}>\ell]=\e^{-\gamma\ell}$
with $\gamma=\xi^{-1}=\kappa_0/D$. The relationship between $p(\x,t|\x_0)$ and $P(\x,\ell,t|\x_0)$ can then be established by noting that
\[p(\x,t|\x_0)d\x=\P[\X_t \in (\x,\x+d\x), \ t < {\mathcal T}|\X_0=\x_0].\]
Given that $\ell_t$ is a nondecreasing process, the condition $t < {\mathcal T}$ is equivalent to the condition $\ell_t <\widehat{\ell}$. This implies that \cite{Grebenkov20}
\begin{eqnarray*}
p(\x,t|\x_0)d\x&=\P[\X_t \in (\x,\x+d\x), \ \ell_t < \widehat{\ell}|\X_0=\x_0]\\
&=\int_0^{\infty} d\ell \ \gamma\e^{-\gamma\ell}\P[\X_t \in (\x,\x+d\x), \ \ell_t < \ell |\X_0=\x_0]\\
&=\int_0^{\infty} d\ell \ \gamma \e^{-\gamma\ell}\int_0^{\ell} d\ell' [P(\x,\ell',t|\x_0)d\x].
\end{eqnarray*}
Using the identity
\[\int_0^{\infty}d\ell \ f(\ell)\int_0^{\ell} d\ell' \ g(\ell')=\int_0^{\infty}d\ell' \ g(\ell')\int_{\ell'}^{\infty} d\ell \ f(\ell)\]
for arbitrary integrable functions $f,g$, it follows that
\begin{equation}
\label{bob}
p(\x,t|\x_0)=\int_0^{\infty} \e^{-\gamma\ell}P(\x,\ell,t|\x_0)d\ell.
\end{equation}
Since the Robin boundary condition maps to an exponential law for the stopping local time $\widehat{\ell}_t$, the probability density $p(\x,t|\x_0)$ can be expressed in terms of the Laplace transform of the propagator $P(\x,\ell,t|\x_0)$ with respect to the local time $\ell$. 
The advantage of this formulation is that one can consider a more general probability distribution $\Psi(\ell) = \P[\ellh>\ell]$ for the stopping local time $\ellh$ such that \cite{Grebenkov19b,Grebenkov20,Grebenkov21}
  \begin{equation}
  \label{oo}
  p(\x,t|\x_0)=\int_0^{\infty} \Psi(\ell)P(\x,\ell,t|\x_0)d\ell \mbox{ for } \x \in \Omega \backslash \calU.
  \end{equation}
This accommodates a wider class of surface reactions where, for example, the reactivity $\kappa(\ell)$ depends on the local time $\ell$ (or the number of surface encounters). The corresponding distribution of the stopping local time $\widehat{\ell}$ would then be
\begin{equation}
\label{kaell}
\Psi(\ell)=\exp\left (-\frac{1}{D}\int_0^{\ell}\kappa(\ell')d\ell'\right ).
\end{equation}

Now suppose that the whole interior target domain $\calU$ acts as a partial absorber rather than the target boundary $\partial \calU$, as shown in Fig. \ref{fig1}(b). That is, the particle can freely enter and exit $\calU$, and is absorbed according to some surface reaction scheme when inside $\calU$. In the case of a constant rate of absorption $k_0$, the BVP for the probability density of particle position can be written down explicitly \cite{Schumm21}. Denoting the probability density by $p(\x,t|\x_0)$ for $\x \in \Omega \backslash \calU$ and by $q(\x,t|\x_0)$ for $\x\in \calU$, we have
\numparts 
\begin{eqnarray}
\label{diffocc1}
\fl 	&\frac{\partial p(\x,t|\x_0)}{\partial t} = D\nabla^2 p(\x,t|\x_0) \mbox{ for }\x\in \Omega\backslash \calU,\quad \nabla p(\x,t|\x_0) \cdot \n=0 \mbox{ for } \x\in\partial \Omega,\\
\fl &\frac{\partial q(\x,t|\x_0)}{\partial t} = D\nabla^2 q(\x,t|\x_0)-k_0 q(\x,t|\x_0) \mbox{ for } \x\in  \calU, 
	\end{eqnarray}
together with the continuity conditions
	\begin{equation}
	\label{diffocc3}
\fl	p(\x,t|\x_0)=q(\x,t|\x_0),\ \nabla p(\x,t|\x_0)\cdot \n =\nabla q(\x,t|\x_0) \cdot \n \ \mbox{ for all } \x \in \partial \calU.
	\end{equation}
	\endnumparts 	
	The initial position of the particle is assumed to be outside the target so that
	\begin{equation}
	p(\x,0|\x_0)=\delta(\x-\x_0),\quad q(\x,0|\x_0)=0.
	\end{equation}
These equations are a direct analog of equations (\ref{diffloc1}) and (\ref{diffloc2}). However, the absorption rate $k_0$ has units of 1/s rather than m/s. As we will show in this paper, the reaction scheme can be generalized along analogous lines to a reactive boundary by replacing the local time $\ell_t$ with the occupation time
\begin{equation}
\label{occ}
A_t=\int_{0}^tI_{\calU}(\X_{\tau})d\tau .
\end{equation}
Here $I_{\calU}(\x)$ denotes the indicator function of the set $\calU\subset \R^d$, that is, $I_{\calU}(\x)=1$ if $\x\in \calU$ and is zero otherwise. Let $P(\x,a,t|\x_0)$ denote the joint probability density or propagator for the pair $(\X_t,A_t)$ and introduce the stopping time
\begin{equation}
\label{TA}
{\mathcal T}=\inf\{t>0:\ A_t >\widehat{A}\},
\end{equation}
where $\widehat{A}$ is a stopping occupation time with probability distribution $\Psi(a)$.
The natural generalization of equation (\ref{oo}) is then
\numparts
\begin{eqnarray}
\label{pP}
p(\x,t|\x_0)&=\int_0^{\infty}\Psi(a) P(\x,a,t|\x_0)da \mbox{ for } \x \in \Omega \backslash \cal U,\\ q(\x,t|\x_0)&=\int_0^{\infty}\Psi(a) Q(\x,a,t|\x_0)da\mbox{ for } \x \in   \cal U.
\label{qQ}
\end{eqnarray}
\endnumparts
We will show that equations (\ref{diffocc1})--(\ref{diffocc3}) are recovered in the case of an exponential law $\Psi(a)=\e^{-k_0 a}$.

 \setcounter{equation}{0}
\section{Derivation of the propagator BVP from a Feynman-Kac formula}

The local time (\ref{loc}) and occupation time (\ref{occ}) are two examples of a Brownian functional. Suppose that $\X_t$ is the position of a Brownian particle at time $t$ with $\X_t \in \R^d$.
A Brownian functional over a fixed time interval $[0,T]$ is defined as a random variable $U_t$ given by \cite{Majumdar05}
\begin{equation}
U_t=\int_0^t F(\X_{\tau})d\tau,
\end{equation}
where $F(\x)$ is some prescribed function or distribution such that $U_t$ has positive support and $\X_0=\x_0$ is fixed. Since $\X_{\tau}$ is a continuous stochastic process, it follows that each realization of a Brownian path will typically yield a different value of $U_t$, which means that $U_t$ is itself a stochastic process. 
Let $P(\x,u,t|\x_0)$ denote the joint probability density or propagator for the pair $(\X_t,U_t)$. It follows that
 \begin{eqnarray}
 \label{A1}
& P(\x,u,t|\x_0)=\bigg \langle \delta\left (u -U_t \right )\bigg \rangle_{\X_0=\x_0}^{\X_t=\x} ,
 \end{eqnarray}
 where expectation is taken with respect to all random paths realized by $\X_{\tau}$ between $\X_0=\x_0$ and $\X_t=\x$. 
  Using a Fourier representation of the Dirac delta function, equation (\ref{A1}) can be rewritten as
 \begin{eqnarray}
 P(\x,u,t|\x_0)=\int_{-\infty}^{\infty} \e^{i\omega u}{\mathcal G}(\x,\omega,t|\x_0)\frac{d\omega}{2\pi},
 \end{eqnarray}
 where $ P(\x,u,t|\x_0)=0$ for $u<0$ and
 \begin{eqnarray}
 {\mathcal G}(\x,\omega,t|\x_0)=\bigg\langle \exp \left ( -i\omega U_t\right )\bigg \rangle_{\X_0=\x_0}^{\X_t=\x}.
 \end{eqnarray}
 We now note that ${\mathcal G}$ is the characteristic functional of $U_t$, whose path-integral representation can be used to derive the following Feynman-Kac equation \cite{Kac49,Majumdar05}:
\begin{eqnarray}
\label{calG}
\fl \frac{\partial \calG(\x,\omega,t|\x_0)}{\partial t}&=D\nabla^2 \calG(\x,\omega,t|\x_0) -i\omega  F(\x) \calG(\x,\omega,t|\x_0) .
\end{eqnarray}
Multiplying equation (\ref{calG}) by $\e^{\i\omega u}$, integrating with respect to $\omega$ and using the identity
\[\frac{\partial }{\partial u} P(\x,u,t|\x_0)\Theta(u) =\int_{-\infty}^{\infty} i\omega  \e^{i\omega u}{\mathcal G}(\x,\omega,t|\x_0)\frac{d\omega}{2\pi},\]
with $\Theta(u)$ the Heaviside function, we obtain the general result
\begin{eqnarray}
\frac{\partial P(\x,u,t|\x_0)}{\partial t}&=D\nabla^2 P(\x,u,t|\x_0)-F(\x) \frac{\partial P}{\partial u}(\x,u,t|\x_0) \nonumber\\
&\quad - \delta(u)F(\x)P(\x,0,t|\x_0),\quad \x\in \R^d  .
\label{calP}
\end{eqnarray}
Equation (\ref{calP}) also holds for diffusion in a bounded domain with totally reflecting boundaries on taking $\X_t$ to be the position of a particle executing reflected Brownian motion.

In the case of the local time (\ref{loc}), the bounded domain is $\Omega\backslash \calU$ and 
\begin{eqnarray}
F(\x)=\lim_{h\rightarrow 0} \frac{D}{h}\Theta(h-\mbox{dist}(\x,\partial \calU))= D\int_{\partial \calU}\delta(\x-\x')d\x'.
\end{eqnarray}
Equation (\ref{calP}) becomes
\begin{eqnarray}
\fl \frac{\partial P(\x,\ell,t|\x_0)}{\partial t}&=D\nabla^2 P(\x,\ell,t|\x_0)\nonumber\\
\fl &\quad -D\int_{\partial \calU}\left (\frac{\partial P}{\partial \ell}(\x',\ell,t|\x_0) +\delta(\ell)P(\x,0,t|\x_0) \right )\delta(\x-\x')d\x',
\end{eqnarray}
which is equivalent to the BVP
\numparts
\begin{eqnarray}
\label{Ploc1}
\fl &\frac{\partial P(\x,\ell,t|\x_0)}{\partial t}=D\nabla^2 P(\x,\ell,t|\x_0),\ \x \in \Omega\backslash \calU, \ \nabla P(\x,\ell,t|\x_0) \cdot \n =0 \mbox{ for } \x\in \partial \Omega,\\
\fl &-D\nabla P(\x,\ell,t|\x_0) \cdot \n= D P(\x,\ell=0,t|\x_0) \ \delta(\ell)  +D\frac{\partial}{\partial \ell} P(\x,\ell,t|\x_0) \mbox{ for }  \x\in \partial \calU.
\label{Ploc2}
\end{eqnarray}
We now note that 
\begin{equation}
\label{Ploc3}
P(\x,\ell=0,t|\x_0)=-\nabla p_{\infty}(\x,t|\x_0)\cdot \n \mbox{ for } \x\in \partial \calU, 
\end{equation}
\endnumparts
where $p_{\infty}$ is the probability density in the case of a totally absorbing target: 
\numparts 
\begin{eqnarray}
\label{pinf}
\fl 	&\frac{\partial p_{\infty}(\x,t|\x_0)}{\partial t} = D\nabla^2 p_{\infty}(\x,t|\x_0), \, \x\in \Omega\backslash \calU,\  \nabla p_{\infty}(\x,t|\x_0) \cdot \n=0 \mbox{ for } \x\in\partial \Omega,\\
\fl &p_{\infty}(\x,t|\x_0)=0,\  \x\in \partial \calU,\ p_{\infty}(\x,0|\x_0)=\delta(\x-\x_0).
	\end{eqnarray}
	\endnumparts 
	The equality (\ref{Ploc3}) can be understood by noting that a constant reactivity is equivalent to a Robin boundary condition, see equation (\ref{bob}). 	In particular, the Robin boundary condition can be rewritten as
\begin{eqnarray}
\fl \nabla p(\x,t|\x_0)\cdot \n&=-\gamma p(\x,t|\x_0)=-\gamma \int_0^{\infty}\e^{-\gamma \ell}P(\x,\ell,t|\x_0)d\ell \mbox{ for } \x \in \partial \calU.
\end{eqnarray}
The result follows from taking the limit $\gamma \rightarrow \infty$ on both sides with $p\rightarrow p_{\infty}$, and noting that $\lim_{\gamma \rightarrow \infty}\gamma \e^{-\gamma \ell}$ is the Dirac delta function on the positive half-line. Note that equations (\ref{Ploc1})--(\ref{Ploc3}) are identical to the BVP derived in Ref. \cite{Grebenkov20} using a different method.

In the case of the occupation time (\ref{occ}), the bounded domain is $\Omega$ and
\begin{equation}
F(\x)=I_{\calU}(\x)=\int_{\calU}\delta(\x-\x')d\x'.
\end{equation}
 Equation (\ref{calP}) now becomes
\begin{eqnarray}
\fl \frac{\partial P(\x,a,t|\x_0)}{\partial t}&=D\nabla^2 P(\x,a,t|\x_0)\nonumber\\
\fl &\quad -\int_{\calU}\left (\frac{\partial P}{\partial a}(\x',a,t|\x_0) +\delta(a)P(\x',0,t|\x_0) \right )\delta(\x-\x')d\x'
\label{Pocc}
\end{eqnarray}
for all $\x \in \Omega$, together with the Neumann boundary condition on $\partial \Omega$. That is,
\numparts
\begin{eqnarray}
\label{Pocc2a}
\fl \frac{\partial P(\x,a,t|\x_0)}{\partial t}&=D\nabla^2 P(\x,a,t|\x_0) \mbox{ for }\x \in \Omega\backslash \calU,\\
\fl \nabla \cdot P(\x,a,t|\x_0)&=0 \mbox{ for } \x \in \partial \Omega, \\
\fl \frac{\partial Q(\x,a,t|\x_0)}{\partial t}&=D\nabla^2 Q(\x,a,t|\x_0) -\left (\frac{\partial Q}{\partial a}(\x,a,t|\x_0) +\delta(a)Q(\x,0,t|\x_0) \right )
\label{Pocc2b}
\end{eqnarray}
for $ \x \in \calU$,
where the propagator within $\calU$ is denoted by $Q$.
We also have the continuity conditions
	\begin{equation}
	\label{Pocc2c}
\fl	P(\x,a,t|\x_0)=Q(\x,a,t|\x_0),\quad \nabla Q(\x,a,t|\x_0)\cdot \n =\nabla P(\x,a,t|\x_0) \cdot \n  
	\end{equation}
	for all $\x \in \partial \calU$.
	\endnumparts
Multiplying both sides of equations (\ref{Pocc2a})--(\ref{Pocc2c}) by $\Psi(a)$ and using (\ref{pP})--(\ref{qQ}) then yields the following generalization of equations (\ref{diffocc1})--(\ref{diffocc3}):
\numparts 
\begin{eqnarray}
\label{new1}
\fl 	&\frac{\partial p(\x,t|\x_0)}{\partial t} = D\nabla^2 p(\x,t|\x_0), \ \x\in \Omega\backslash \calU,\quad \nabla p(\x,t|\x_0) \cdot \n=0 \mbox{ for } \x\in\partial \Omega,\\
\label{new2}
\fl &\frac{\partial q(\x,t|\x_0)}{\partial t} = D\nabla^2 q(\x,t|\x_0)-\int_0^{\infty} \psi(a) Q(\x,a,t|\x_0)da \mbox{ for } \x\in  \calU,\\
 \fl & p(\x,0|\x_0)=\delta(\x-\x_0),\quad q(\x,0|\x_0)=0,
	\end{eqnarray}
together with the continuity conditions
	\begin{equation}
	\label{new3}
\fl	p(\x,t|\x_0)=q(\x,t|\x_0),\ \nabla p(\x,t|\x_0)\cdot \n =\nabla q(\x,t|\x_0) \cdot \n \ \mbox{ for all } \x \in \partial \calU.
	\end{equation}
	\endnumparts 	
	Analogous to the local time, we have set $\psi(a)=-\Psi'(a)$. Clearly equations (\ref{diffocc1})--(\ref{diffocc3}) are recovered in the case of a constant reactiion rate, $\psi(a)=k_0 \e^{-k_0a}$.

Having solved the appropriate BVP for the propagator $P$, we can then determine the probability density $p(\x,t|\x_0)$ according to equation (\ref{oo}) or (\ref{occ}),
and use this to investigate the statistics of the absorption process. A typical quantity of interest is the mean first passage time (MFPT) for absorption. First, consider absorption at the target boundary $\partial \calU$. The survival probability that the particle hasn't been absorbed by the target in the time interval $[0,t]$, having started at $\x_0$, is defined according to
\begin{equation}
\label{S1}
S(\x_0,t)=\int_{\Omega\backslash \calU}p(\x,t|\x_0)d\x.
\end{equation}
Differentiating both sides of this equation with respect to $t$ and using the diffusion equation implies that
\begin{eqnarray}
\frac{\partial S(\x_0,t)}{\partial t}&=D\int_{\Omega\backslash \calU_a}\nabla\cdot \nabla p(\x,t|\x_0)d\x=D \int_{ \partial \calU}\nabla p(\x,t|\x_0)\cdot \n d\sigma\nonumber \\
& =-J(\x_0,t),
\label{Q2}
\end{eqnarray}
where $d\sigma$ is the surface measure and $J(\x_0,t)$ is the probability flux into the target at time $t$. We have used the divergence theorem and the Neumann boundary condition on $\partial \Omega$.
Laplace transforming equation (\ref{Q2}) and noting that $S(\x_0,0)=1$ gives
\begin{equation}
\label{QL}
s\widetilde{S}(\x_0,s)-1=- \widetilde{J}(\x_0,s).
\end{equation}
Taking the limit $s\rightarrow 0$ on both sides and noting that $S(\x_0,t)\rightarrow 0$ as $t\rightarrow \infty$, we see that
$\lim_{s\rightarrow 0}\widetilde{J}(\x_0,s)=1$.
The probability density of the stopping time $\calT$, equation (\ref{Tell}), is given by $-\partial S/\partial t$ so that the MFPT is 
\begin{eqnarray}
\label{MFPT1}
T(\x_0)&=-\int_0^{\infty}t\frac{\partial S(\x_0,t)}{\partial t}dt =\int_0^{\infty} S(\x_0,t)dt \nonumber \\
&= \widetilde{S}(\x_0,s)=\lim_{s\rightarrow 0}\frac{1-\widetilde{J}(\x_0,s)}{s}=-\left .\frac{\partial}{\partial s}\widetilde{J}(\x_0,s)\right |_{s=0}.
\end{eqnarray}
We have used integration by parts. Note that the Laplace transformed flux can be expressed directly in terms of the propagator using the boundary condition (\ref{Ploc2}). Multiplying both sides of the latter by $\Psi(\ell)$ and integrating by parts with respect to $\ell$ shows that
\begin{eqnarray}
-D\nabla p(\x,t|\x_0) \cdot \n=D\int_0^{\infty}\psi(\ell) P(\x,\ell,t|\x_0)d\ell \mbox{ for } \x\in \partial \calU,
\end{eqnarray}
with
\begin{equation}
\psi(\ell)=-\frac{d\Psi(\ell)}{d\ell}, \quad \widetilde{\psi}(q)=1-q\widetilde{\Psi}(q).
\end{equation}
We have used equation (\ref{Ploc3}) and the identity $\Psi(0)=1$. Integrating with respect to points on the boundary and Laplace transforming gives
\begin{equation}
\label{J}
\widetilde{J}(\x_0,s)=D\int_0^{\infty}\psi(\ell)\left [ \int_{\partial \calU}\PP(\x,\ell,s|\x_0)d\sigma \right ]d\ell.
\end{equation}

In the case of an absorbing target interior, the survival probability is
\begin{equation}
\label{Socc}
\fl S(\x_0,t)=\int_{\Omega}p(\x,t|\x_0)d\x= \int_{\Omega\backslash \calU}  p(\x,t|\x_0)d\x+\int_{ \calU}  q(\x,t|\x_0)d\x.
\end{equation}
Differentiating with respect to $t$ and using equations (\ref{new1}) and (\ref{new2}) gives
\begin{eqnarray}
\frac{\partial S(\x_0,t)}{\partial t}&=D\int_{\Omega\backslash \calU} \nabla^2 p(\x,t|\x_0)d\x+D \int_{  \calU}\nabla^2 q(\x,t|\x_0)d\x\nonumber \\
&\quad -\int_{\calU} \int_0^{\infty} \psi(a) Q(\x,a,t|\x_0)da\ d\x.
\label{Qocc}
\end{eqnarray}
Applying the divergence theorem to the first two integrals on the right-hand side, imposing the Neumann boundary condition on $\partial \Omega$ and flux continuity at $\partial \calU$ shows that these two integrals cancel. The result is then
\begin{eqnarray}
\frac{\partial S(\x_0,t)}{\partial t}&=-\int_{\calU} \int_0^{\infty} \psi(a) Q(\x,a,t|\x_0)da\ d\x =-K(\x_0,t),
\end{eqnarray}
where $K(\x_0,t)$ is the probability flux due to absorption within the target domain $\calU$. Using a similar argument to the previous case, we find that the MFPT is
\begin{equation}
\label{MFPT2}
T(\x_0)=\lim_{s\rightarrow 0}\frac{1-\widetilde{K}(\x_0,s)}{s}=-\left .\frac{\partial}{\partial s}\widetilde{K}(\x_0,s)\right |_{s=0},
\end{equation}
with
\begin{equation}
\label{Ktil}
\widetilde{K}(\x_0,s)=\int_0^{\infty}\psi(a) \left [\int_{\calU}\widetilde{Q}(\x,a,s|\x_0)d\x \right ]da.
\end{equation}

\section{MFPT for a spherical target}

In order to illustrate the above theory, consider a spherical domain $\Omega =\{\x\in \R^d\,|\, 0\leq  |\x| <R_2\}$ and a spherical target of radius $R_1$ at the center of $\Omega$ with $R_1<R_2$:
\[\calU=\{\x\in \R^d \, |\, 0\leq  |\x|<R_1\},\quad \partial \calU=\{\x\in \R^d\, |\, |\x|=R_1\}.\]
Following \cite{Redner01}, the initial position of the particle is randomly chosen from the surface of the sphere of radius $\rho_0$, $R_1<\rho_0<R_2$.  That is,
\begin{eqnarray}
    p({\bf x}, 0|{\bf x}_0) = \frac{1}{\Omega_d \rho_0^{d - 1}} \delta(\rho - \rho_0),
\end{eqnarray}
where $\rho= \|{\bf x}\|$ and $\Omega_d$ is the surface area of a unit sphere in $\mathbb{R}^d$. This allows us to exploit spherical symmetry such that $p=p(\rho,t|\rho_0)$ and $P=P(\rho,\ell,t|\rho_0)$. Note that the same spherical shell configuration is considered in Ref. \cite{Grebenkov20} in the case of reaction at the target boundary. However, the propagator is obtained using a spectral decomposition of the Dirichlet-Neumann operator rather than by solving the propagator BVP directly. Moreover, the MFPT is not considered.

\subsection{Absorption at the target boundary}

Laplace transforming equations (\ref{Ploc1})-(\ref{Ploc3}) and introducing spherical polar coordinates gives
  \numparts
\begin{eqnarray}
\label{spha}
 \fl   &D\frac{\partial^2\PP}{\partial \rho^2} + D\frac{d - 1}{\rho}\frac{\partial \PP}{\partial \rho} -s\PP(\rho,\ell, s|\rho_0) = -\frac{1}{\Omega_d \rho_0^{d - 1}} \delta(\rho - \rho_0)\delta(\ell), \ R_1<\rho <R_2,\\ 
 \fl & \left . \frac{\partial }{\partial \rho}\PP(\rho,\ell,s|\rho_0)\right |_{\rho=R_2}=0,
  \label{sphb}\\
\fl &\frac{\partial }{\partial \rho}\PP(\rho,\ell,s|\rho_0) =  \PP(\rho,\ell=0,s|\rho_0) \ \delta(\ell)  +\frac{\partial}{\partial \ell} \PP(\rho,\ell,s|\rho_0),\  \rho=R_1.
 \label{sphc}
\end{eqnarray}
\endnumparts
 Equations of the form (\ref{spha}) can be solved in terms of modified Bessel functions \cite{Redner01}. The general solution is
  \begin{eqnarray}
\label{qir}
 \fl    \PP(\rho,\ell, s|\rho_0) = B(\ell) \rho^\nu I_\nu(\alpha \rho)  + C(\ell)\rho^\nu K_\nu(\alpha \rho) + G(\rho, s| \rho_0)\delta(\ell), \ \rho \in (R_1, R_2),
\end{eqnarray}
with $\nu = 1 - d/2$ and $\alpha=\sqrt{s/D}$. In addition, $I_{\nu}$ and $K_{\nu}$ are modified Bessel functions of the first and second kind, respectively. 
The first two terms on the right-hand side of equation (\ref{qir}) are the solutions to the homogeneous version of equation (\ref{spha}) and $G$ is the Green's function satisfying
  \numparts
\begin{eqnarray}
\label{Ga}
 \fl   &D\frac{\partial^2G}{\partial \rho^2} + D\frac{d - 1}{\rho}\frac{\partial G}{\partial \rho} -sG  = -\frac{1}{\Omega_d \rho_0^{d - 1}} \delta(\rho - \rho_0), \ R_1<\rho <R_2,\\ 
 \fl & G(R_1,s|\rho_0)=0,\quad \left .\frac{\partial }{\partial \rho}G(\rho,s|\rho_0)\right |_{\rho=R_2}=0.
 \label{Gb}
\end{eqnarray}
\endnumparts
The latter is given by \cite{Redner01}
\begin{eqnarray}
\label{GG}
  G(\rho, s| \rho_0) = \frac{ (\rho\rho_0)^\nu }{D\Omega_d}\frac{C_{\nu}(\rho_{<},R_1;s)D_{\nu}(\rho_{>},R_2;s)}{D_{\nu}(R_1,R_2;s)},
\end{eqnarray}
where $\rho_< = \min{(\rho, \rho_0)}$, $\rho_> = \max{(\rho, \rho_0)}$, and
\numparts
\begin{eqnarray}
 C_{\nu}(a,b;s)&=I_{\nu}(\alpha a)K_{\nu}(\alpha b)-I_{\nu}(\alpha b)K_{\nu}(\alpha a),\\
  D_{\nu}(a,b;s)&=I_{\nu}(\alpha a)K_{\nu-1}(\alpha b)+I_{\nu-1}(\alpha b)K_{\nu}(\alpha a).
\end{eqnarray}
\endnumparts

The unknown coefficients $B(\ell)$ and $C(\ell)$ are determined from the boundary conditions (\ref{sphb}) and (\ref{sphc}). In order to simplify the notation, we set
\begin{equation}
F_I(\rho)=\rho^\nu I_\nu(\alpha \rho),\quad F_K(\rho)=\rho^\nu K_\nu(\alpha \rho).
\end{equation}
\numparts
Note that $F_{I,K}$ are also functions of $\alpha$.
Equation (\ref{sphb}) becomes 
\begin{equation}
\label{CB1}
B(\ell) F_I'(R_2)+C(\ell)F_K'(R_2)=0.
\end{equation}
Since  $G(\rho,s|\rho_0)\equiv p_{\infty}(\rho,s|\rho_0)$, equation (\ref{sphc}) implies that 
\begin{eqnarray}
\label{CB2}
  &\frac{dB(\ell)}{d\ell} F_I(R_1) +\frac{dC(\ell)}{d\ell} F_K(R_1) =B(\ell)F_I'(R_1)+C(\ell) F'_K(R_1), 
\end{eqnarray}
and
\begin{eqnarray}
\fl B(0)F_I(R_1) +C(0)F_K(R_1)&=\left . \frac{d}{d\rho}G(\rho,s|\rho_0) \right |_{\rho=R_1}\equiv \frac{1}{D\Omega_d}\left (\frac{\rho_0}{R_1}\right )^{\nu} \frac{D_{\nu}(\rho_0,R_2)}{D_{\nu}(R_1,R_2)}.
\label{CB3}
\end{eqnarray}
\endnumparts
Equation (\ref{CB1}) shows that
\begin{equation}
B(\ell)=-\frac{F_K'(R_2)}{F_I'(R_2)}C(\ell).
\end{equation}
Substituting into equation (\ref{CB2}) and rearranging yields the following differential equation for $C(\ell)$: 
\begin{eqnarray}
\label{C}
\frac{dC(\ell)}{d\ell}+\Lambda_{\alpha}(R_1,R_2)C(\ell)=0,
\end{eqnarray}
with
\begin{eqnarray}
\label{GamR}
\fl \Lambda_{\alpha}(R_1,R_2)=-\left (F_K(R_1)-F_I(R_1)\frac{F_K'(R_2)}{F_I'(R_2)}\right )^{-1}\left (F_K'(R_1)-F_I'(R_1)\frac{F_K'(R_2)}{F_I'(R_2)}\right ).
\end{eqnarray}
Equation (\ref{C}) has the solution
\begin{equation}
C(\ell)=C(0)\e^{-\Lambda_{\alpha}(R_1,R_2)\ell},
\end{equation}
with
\begin{equation}
C(0)= \left (F_K(R_1)-F_I(R_1)\frac{F_K'(R_2)}{F_I'(R_2)}\right )^{-1}\left . \frac{d}{d\rho}G(\rho,s|\rho_0) \right |_{\rho=R_1}.
\end{equation}

Combining our various results shows that at the surface of the target ($\rho =R_1)$,
\begin{eqnarray}
\label{qir2}
 \PP(R_1,\ell, s|\rho_0) &= B(\ell) F_I(R_1) + C(\ell)F_K(R_1) \nonumber \\
 &= \left (F_K(R_1)-F_I(R_1)\frac{F_K'(R_2)}{F_I'(R_2)}\right )C(0)\e^{-\Lambda_{\alpha}(R_1,R_2)\ell}\nonumber \\
 &=\left . \frac{d}{d\rho}G(\rho,s|\rho_0) \right |_{\rho=R_1}\e^{-\Lambda_{\alpha}(R_1,R_2)\ell}.
\end{eqnarray}
Substituting into equation (\ref{J}) shows that the Laplace-transformed flux into the spherical target is
\begin{eqnarray}
\widetilde{J}(\rho_0,s)&=D\Omega_dR_1^{d-1} \left . \frac{d}{d\rho}G(\rho,s|\rho_0) \right |_{\rho=R_1} \int_0^{\infty}\psi(\ell) \e^{-\Lambda_{\alpha}(R_1,R_2)\ell} d\ell\nonumber \\
&=\widetilde{J}_{\infty}(\rho_0,s)\widetilde{\psi}(\Lambda_{\alpha}(R_1,R_2)),
\label{J2}
\end{eqnarray}
where
\begin{equation}
\widetilde{J}_{\infty}(\rho_0,s)=\left (\frac{\rho_0}{R_1}\right )^{\nu} \frac{D_{\nu}(\rho_0,R_2)}{D_{\nu}(R_1,R_2)}
\end{equation}
is the corresponding flux into a totally absorbing target. Finally, differentiating equation (\ref{J2}) with respect to $s$ and using equation (\ref{MFPT1}), we obtain the result
\begin{eqnarray}
T(\rho_0)&=-\left .\frac{\partial}{\partial s}\widetilde{J}(\rho_0,s)\right |_{s=0}\nonumber\\
&=T_{\infty}(\rho_0)-\lim_{s\rightarrow 0} \frac{1}{2\sqrt{sD}} \frac{d}{d\alpha}\widetilde{\psi}(\Lambda_{\alpha}(R_1,R_2))\nonumber\\
&=T_{\infty}(\rho_0)-\widetilde{\psi}'(0)\lim_{s\rightarrow 0} \frac{1}{2\sqrt{sD}}\frac{d}{d\alpha}\Lambda_{\alpha}(R_1,R_2),
\label{Tfin}
\end{eqnarray}
where
\begin{equation}
T_{\infty}(\rho_0)=-\left .\frac{\partial}{\partial s}\widetilde{J}_{\infty}(\rho_0,s)\right |_{s=0}
\end{equation}
is the MFPT in the case of a totally absorbing target. It immediately follows that if a surface reaction involves a stopping local time distribution with $\widetilde{\psi}'(0)=-\infty$, then the MFPT $T(\rho_0)$ blows up, indicating that the target is not sufficiently absorbing. In other words, for finite $T(\rho_0)$ the stopping local time density $\psi(\ell)$ must have a finite first moment, since
\begin{equation}
\widetilde{\psi}'(0)=-\int_0^{\infty}\ell\psi(\ell)d\ell.
\end{equation}
Finally, as $R_2\rightarrow R_1$, we have $\Lambda_{\alpha}(R_1,R_2) \rightarrow 0$ and $\widetilde{J}_{\infty}(\rho_0,s)\rightarrow 1$, which means that $T(\rho_0)\rightarrow 0$. This reflects the fact that the particle never spends any time away from the target boundary and, hence, the survival time is identically zero.

\begin{figure}[b!]
\raggedleft
\includegraphics[width=10cm]{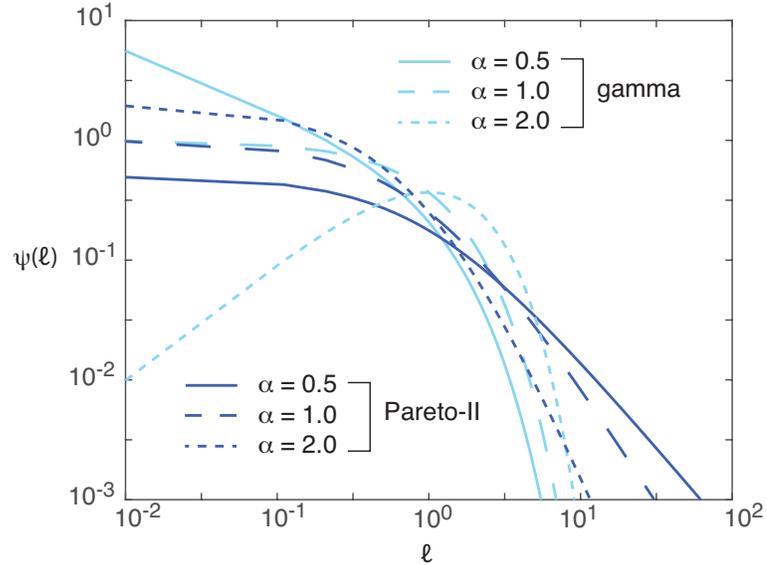} 
\caption{Plots of the probability density $\psi(\ell)$ as a function of the stopping local time for the gamma and Pareto-II models. We also set $\gamma=\kappa_0/D=1$.}
\label{fig2}
\end{figure}

We now consider two examples of surface reactions whose stopping local time distributions are given by the gamma distribution and the Pareto-II (Lomax) distribution, respectively, see Fig. \ref{fig2}. A more comprehensive list of models is given in Table 1 of Ref. \cite{Grebenkov20}. In both cases we take $\gamma=\kappa_0/D$, where $\kappa_0$ is some reference reactivity. 
\medskip

\noindent  {\em (a) Gamma distribution.}
First consider the gamma distribution and its equivalent encounter-dependent reactivities $\kappa(\ell)$, see equation (\ref{kaell}): 
\begin{equation}
\psi(\ell)=\frac{\gamma(\gamma \ell)^{\mu-1}\e^{-\gamma \ell}}{\Gamma(\mu)},\quad \kappa(\ell)=\kappa_0 \frac{ (\gamma \ell)^{\mu-1}\e^{-\gamma \ell}}{\Gamma(\mu,\gamma \ell)} ,\ \mu >0,
\end{equation}
where $\Gamma(\mu)$ is the gamma function and $\Gamma(\mu,z)$ is the upper incomplete gamma function:
\begin{equation}
\Gamma(\mu)=\int_0^{\infty}\e^{-t}t^{\mu-1}dt,\quad \Gamma(\mu,z)=\int_z^{\infty}\e^{-t}t^{\mu-1}dt,\ \mu >0.
\end{equation}
Note that if $\mu=1$ then we obtain the exponential distribution (constant reactivity)
\begin{equation}
\psi(\ell)=\gamma \e^{-\gamma \ell},\quad \kappa(\ell) =\kappa_0.
\end{equation}
The corresponding Laplace transforms are
\begin{equation}
\widetilde{\psi}(q)=\left (\frac{\gamma}{\gamma+q}\right )^{\mu},\quad \widetilde{\psi}'(q)=-\mu \left (\frac{\gamma}{\gamma+q}\right )^{\mu}\frac{1}{\gamma+q},
\end{equation}
It can be seen that $\widetilde{\psi}(0)=1$ and $\widetilde{\psi}'(0)=-\mu/\gamma<-\infty$.
\medskip

\noindent  {\em (b) Pareto-II (Lomax) distribution.}
As a second example, consider the Pareto-II (Lomax) distribution
\begin{equation}
\psi(\ell)=\frac{\gamma \mu}{(1+\gamma \ell)^{1+\mu}},\quad \kappa(\ell)=\kappa_0 \frac{\mu}{1+\gamma \ell} ,\quad \mu >0,
\end{equation}
with
\numparts
\begin{eqnarray}
\fl \widetilde{\psi}(q)&=\mu\left (\frac{q}{\gamma}\right )^{\mu}\e^{q/\gamma}\Gamma(-\mu,q/\gamma),\\
\fl \widetilde{\psi}'(q)&=\mu\left (\frac{q}{\gamma}\right )^{\mu}\e^{q/\gamma}\left (\left [\frac{\mu}{q}+\frac{1}{\gamma}\right ]\Gamma(-\mu,q/\gamma)+\partial_{q} \Gamma(-\mu,q/\gamma)\right ).
\end{eqnarray}
\endnumparts 
Using the identity
\begin{equation}
\Gamma(1-\mu,z)=-\mu \Gamma(-\mu,z) +z^{-\mu}\e^{-z},
\end{equation}
it can be checked that $\widetilde{\psi}(0)=1$, whereas $\widetilde{\psi}'(0)$ is only finite if $\mu>1$. In the latter case
\begin{equation}
-\widetilde{\psi}'(0)=\E[\ell]= \frac{\Gamma(\mu-1)\Gamma(2)}{\gamma \Gamma(\mu)}=\frac{1}{\gamma (\mu-1)}.
\end{equation}
The blow up of the moments when $\mu<1$ reflects the fact that the Paretto-II distribution has a long tail.

\begin{figure}[b!]
  \raggedleft
  \includegraphics[width=10cm]{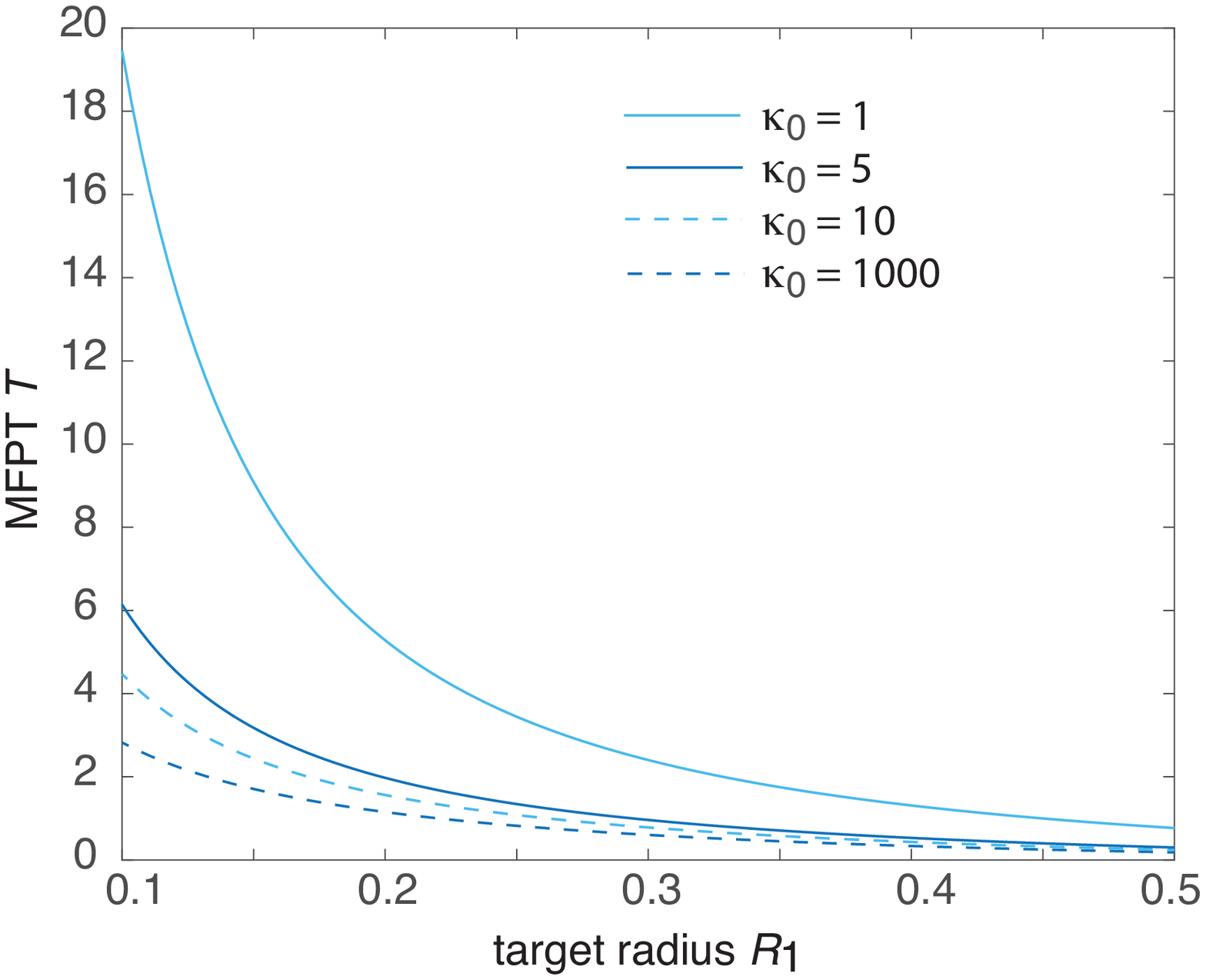}
  \caption{3D spherical target. Plot of MFPT $T(\rho_0)$ as a function of target radius $R_1$ for various absorption rates $\kappa_0$ in the case of the gamma distribution with $\mu=0.5$. Other parameter values are $R_2=1$, $\rho_0=0.75$ and $D=1$.}
  \label{fig3}
\end{figure}

\begin{figure}[t!]
  \raggedleft
  \includegraphics[width=10cm]{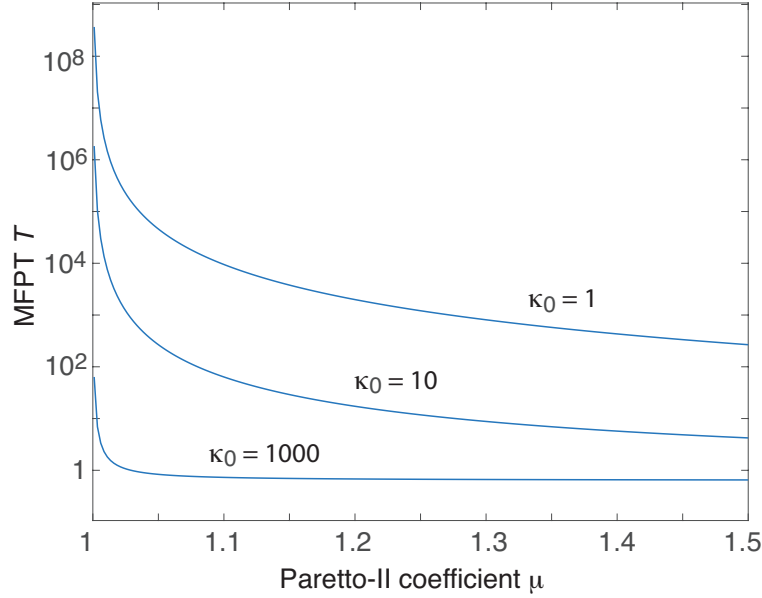}
  \caption{3D spherical target. Plot of MFPT $T(\rho_0)$ as a function of the coefficient $\mu$ for various absorption rates $\kappa_0$ in the case of the Paretto-II distribution. Other parameter values are $R_1=0.25,R_2=1$, $\rho_0=0.75$ and $D=1$.}
  \label{fig4}
\end{figure}

For the sake of illustration, consider the case $d=3$ for which $\nu = -1/2$ and
\begin{eqnarray}
  & I_{-1/2}(z) = \sqrt{\frac{2}{\pi z}}\cosh{(z)},\quad  I_{1/2}(z) = \sqrt{\frac{2}{\pi z}}\sinh{(z)},\\
&K_{\pm 1/2}(z) = \sqrt{\frac{\pi}{2 z}}\e^{-z},\quad K_{\pm 3/2}(z) = \sqrt{\frac{\pi}{2 z}}\frac{\e^{-z}}{1+z}.
    \end{eqnarray}
The coefficient $\Lambda_{\alpha}$ takes the form
\begin{equation}
\label{alpG2}
\fl \Lambda_{\alpha}(R_1,R_2)=\frac{\left [\frac{1}{R_1}+\alpha\right ]\left [1+\frac{1}{\Lambda_{\alpha}(R_2)}\right ] -\frac{1}{\Lambda_{\alpha}(R_2)}\left [\alpha-\frac{1}{R_1}\right ]\e^{2\alpha R_1}}{\left [1+\frac{1}{\Lambda_{\alpha}(R_2)}\right ] +\frac{1}{\Lambda_{\alpha}(R_2)}\e^{2\alpha R_1}},
\end{equation}
with
\begin{equation}
\label{alpG1}
\Lambda_{\alpha}(R_2)= -\pi \frac{F_I'(R_2)}{F_K'(R_2)}=\frac{\alpha R_2-1}{\alpha R_2+1}\e^{2\alpha R_2}-1.
\end{equation}
Note that if $\alpha >0$ then $\Lambda(R_2)\rightarrow \infty$ as $R_2\rightarrow \infty$ (unbounded domain $\Omega = \R^3$), and
\begin{equation}
\Lambda_{\alpha}(R_1,\infty)=\frac{1}{R_1}+\alpha.
\end{equation}
Taylor expanding with respect to $\alpha$, we find
\begin{equation}
\label{aexp}
\fl \Lambda_{\alpha}(R_2)\sim -2+\frac{2}{3}\alpha^3R_2^3+O(\alpha^4),\quad \Lambda_{\alpha}(R_1,R_2)\sim \frac{\alpha^2}{3R_1^2}(R_2^3-R_1^3).
\end{equation}
It follows that for finite $R_2$ we have
\begin{equation}
\lim_{\alpha \rightarrow 0}\Lambda_{\alpha}(R_1,R_2)=0,
\end{equation}
which ensures that
\begin{equation}
\lim_{s\rightarrow 0}\widetilde{J}(\rho_0,s)=\widetilde{\psi}(0)\lim_{s\rightarrow 0}\widetilde{J}_{\infty}(\rho_0,s)=1.
\end{equation}
Moreover, the MFPT (\ref{Tfin}) becomes
\begin{eqnarray}
T(\rho_0)
&=T_{\infty}(\rho_0)-\widetilde{\psi}'(0)\frac{R_2^3-R_1^3}{3DR_1^2}.
\label{Tfin3D}
\end{eqnarray}
In particular, for the gamma distribution $(\mu >0)$ and the Paretto-II distribution ($\mu >1$)
\begin{eqnarray}
\fl T_{\rm gam}(\rho_0)
&=T_{\infty}(\rho_0)+\mu \frac{R_2^3-R_1^3}{3\kappa_0R_1^2}, \quad T_{\rm Par}(\rho_0)
&=T_{\infty}(\rho_0)+\frac{1}{\mu-1} \frac{R_2^3-R_1^3}{3\kappa_0R_1^2}.
\end{eqnarray}
Clearly 
\begin{equation}
\left . T_{\rm gam}(\rho_0)\right |_{\mu =\mu_0}=\left . T_{\rm Par}(\rho_0)\right |_{\mu=1+1/\mu_0}.
\end{equation}

In Fig. \ref{fig3} we show sample plots of $T_{\rm gam}(\rho_0)$ as a function of the inner radius $R_1$ and different absorption rates $\kappa_0$. We take $\rho_0=0.75$ and $R_2=1$. Clearly, as $R_1$ increases, the size of the target grows and the MFPT decreases. In addition, as $\kappa_0\rightarrow \infty$, we have $T_{\rm gam}(\rho_0)\rightarrow T_{\infty}(\rho_0)$. In Fig. 4 we illustrate the blow up of the MFPT as $\mu \rightarrow 1^+$ in the Paretto-II model.

\begin{figure}[t!]
  \raggedleft
  \includegraphics[width=10cm]{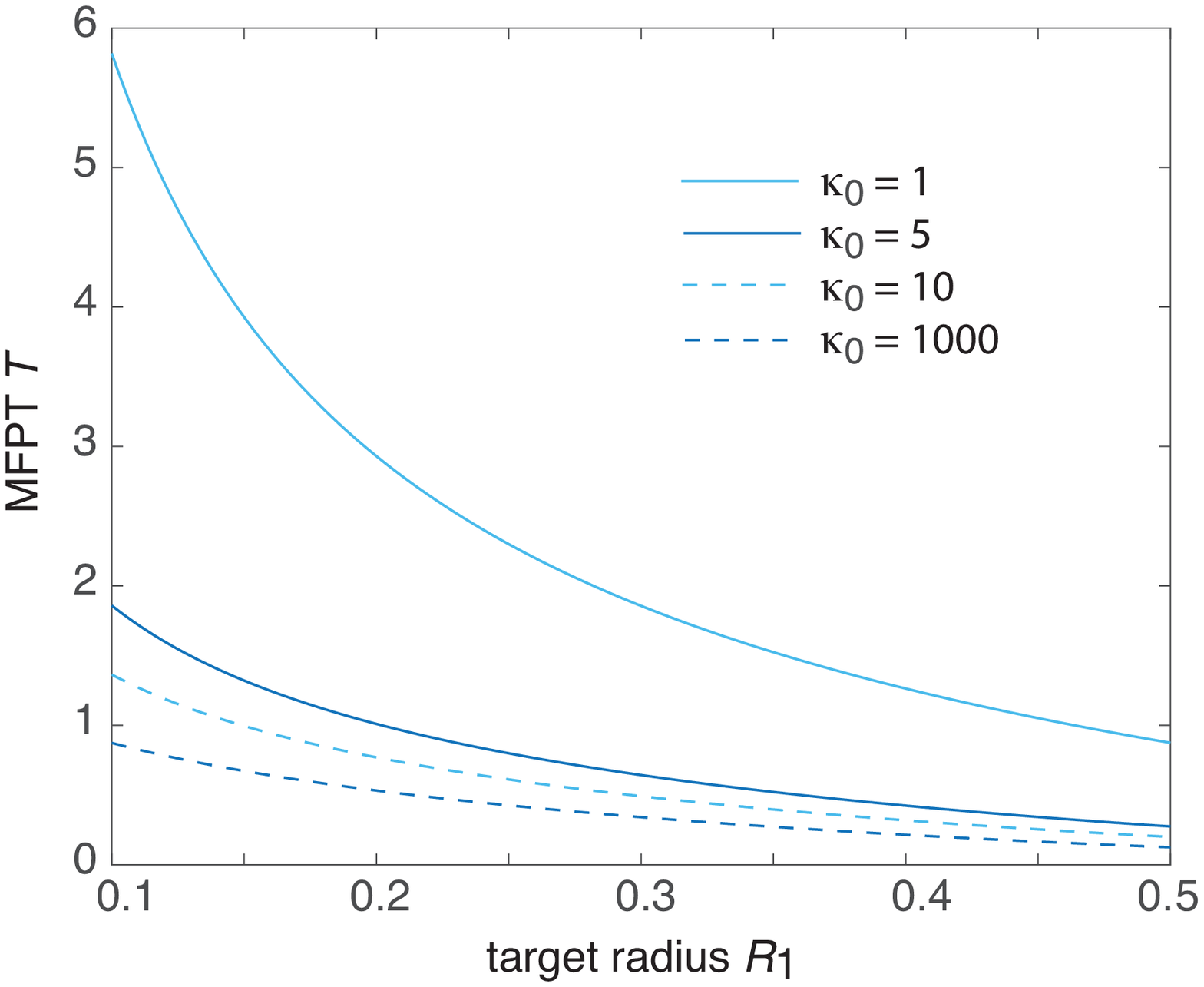}
  \caption{2D spherical target. Plot of MFPT $T(\rho_0)$ as a function of target radius $R_1$ for various absorption rates $\kappa_0$ in the case of the gamma distribution with $\mu=0.5$. Other parameter values are $R_2=1$, $\rho_0=0.75$ and $D=1$.}
  \label{fig5}
\end{figure}

Turning to the case of a 2D sphere ($d=2$), the coefficient $\Lambda_{\alpha}(R_1,R_2)$ takes the form
\begin{eqnarray}
\Lambda_{\alpha}(R_1,R_2)= \frac{\alpha K_{-1}(\alpha R_1)-\alpha I_{-1}(\alpha R_1)\frac{\displaystyle K_{-1}(\alpha R_2)}{\displaystyle I_{-1}(\alpha R_2)}}{K_{0}(\alpha R_1)+I_{0}(\alpha R_1)\frac{\displaystyle K_{-1}(\alpha R_2)}{\displaystyle I_{-1}(\alpha R_2)}}.
\end{eqnarray}
We have used the Bessel function identities
\begin{eqnarray}
I_{\nu}'(z)&=-\frac{\nu}{r}I_{\nu}(z)+I_{\nu-1}(z),\quad K_{\nu}'(z)=-\frac{\nu}{z}K_{\nu}(z)-K_{\nu-1}(z).
\end{eqnarray}
Given the asymptotic expansions
\begin{eqnarray}
\fl I_0(z)\sim 1+\frac{z^2}{4},\ I_{-1}(z)=I_1(z)\sim \frac{z}{2}, \quad K_0(z)\sim -\ln z,\ K_{-1}(z)=K_1(z)\sim \frac{1}{z},
\end{eqnarray}
it follows that for small $\alpha$,
\begin{eqnarray}
\fl \Lambda_{\alpha}(R_1,R_2)\sim \frac{\alpha (1-R_1^2/R_2^2)}{-\alpha R_1\ln (\alpha R_1) +2\alpha R_1 (1+(\alpha R_1)^2/4)/(\alpha R_2)^2}\sim \frac{\alpha^2}{2R_1}(R_2^2-R_1^2).
\end{eqnarray}
The 2D analog of equation is thus
\begin{eqnarray}
T(\rho_0)
&=T_{\infty}(\rho_0)-\widetilde{\psi}'(0)\frac{R_2^2-R_1^2}{R_1}.
\label{Tfin2D}
\end{eqnarray}
Sample plots are shown in Fig. \ref{fig5}

\subsection{Absorption within the target interior}

Laplace transforming equations (\ref{Pocc2a})-(\ref{Pocc2c}) and rewriting in terms of spherical polar coordinates gives
  \numparts
\begin{eqnarray}
\label{ospha}
 \fl   &D\frac{\partial^2\PP}{\partial \rho^2} + D\frac{d - 1}{\rho}\frac{\partial \PP}{\partial \rho} -s\PP(\rho,a, s|\rho_0) = -\frac{1}{\Omega_d \rho_0^{d - 1}} \delta(\rho - \rho_0)\delta(a), \ R_1<\rho <R_2,\\ 
 \fl & \left . \frac{\partial }{\partial \rho}\PP(\rho,a,s|\rho_0)\right |_{\rho=R_2}=0,
  \label{osphb}\\
\fl   &D\frac{\partial^2\QQ}{\partial \rho^2} + D\frac{d - 1}{\rho}\frac{\partial \QQ}{\partial \rho} -s\QQ(\rho,a, s|\rho_0)  =\frac{\partial \QQ}{\partial a}(\rho,a,s|\rho_0) +\delta(a)\QQ(\rho,0,t|\rho_0) 
 \label{osphc}
\end{eqnarray}
for $0<\rho <R_1$.
\endnumparts
We also have the continuity conditions
	\begin{equation}
	\label{osphd}
\fl	\PP(R_1,a,s|\rho_0)=\QQ(R_1,a,t|\rho_0),\ \nabla \PP(R_1,a,t|\rho_0)\cdot \n =\nabla \QQ(R_1,a,t|\rho_0) \cdot \n .
	\end{equation}
In order to solve the above system of equations, it is convenient to Laplace transform with respect to $a$ by setting
\begin{equation}
\fl {\mathcal P}(\rho,z, s|\rho_0) =\int_0^{\infty}\e^{-az} \PP(\rho,a, s|\rho_0) da,\  {\mathcal Q}(\rho,z, s|\rho_0) =\int_0^{\infty}\e^{-az} \QQ(\rho,a, s|\rho_0) da.
\end{equation}
This gives
  \numparts
\begin{eqnarray}
\label{LTospha}
 \fl   &D\frac{\partial^2\calP}{\partial \rho^2} + D\frac{d - 1}{\rho}\frac{\partial \calP}{\partial \rho} -s\calP(\rho,z, s|\rho_0) = -\frac{1}{\Omega_d \rho_0^{d - 1}} \delta(\rho - \rho_0), \ R_1<\rho <R_2,\\ 
 \fl & \left . \frac{\partial }{\partial \rho}\calP(\rho,z,s|\rho_0)\right |_{\rho=R_2}=0,
  \label{LTosphb}\\
\fl   &D\frac{\partial^2\calQ}{\partial \rho^2} + D\frac{d - 1}{\rho}\frac{\partial \calQ}{\partial \rho} -(s+z)\calQ(\rho,z, s|\rho_0) =0,\quad 0<\rho < R_1,
 \label{LTosphc}
\end{eqnarray}
\endnumparts
together with the corresponding continuity conditions in Laplace space.

 The general solution of equation (\ref{LTospha}) is identical in form to equation (\ref{qir}),
  \begin{eqnarray}\label{cP0}
 \fl    \calP(\rho,z, s|\rho_0) = {B}(z) \rho^\nu I_\nu(\alpha \rho)  + {C}(z)\rho^\nu K_\nu(\alpha \rho) + G(\rho, s| \rho_0) \ R_1<\rho <R_2.
 \end{eqnarray}
Similarly, the homogeneous equation (\ref{LTosphc}) has the solution
 \begin{eqnarray}
 \label{cQ0}
    \calQ(\rho,z, s|\rho_0) = \widehat{B}(z) \rho^\nu I_\nu(\beta \rho)  + \widehat{C}(z)\rho^\nu K_\nu(\beta\rho), \ 0<\rho < R_1,
\end{eqnarray}
with
\begin{equation}
\beta =\sqrt{\frac{s+z}{D}}.
\end{equation}
There are four unknown coefficients but only one boundary condition (\ref{LTosphb}) and two continuity conditions. The fourth condition is obtained by requiring that the solution remains finite at $\rho=0$. The details of the latter will depend on the dimension $d$. We will focus on the 3D case for which equations (\ref{cP0}) and (\ref{cQ0}) become
 \begin{eqnarray}\label{cP}
 \fl    \calP(\rho,z, s|\rho_0) = {B}(z) \sqrt{\frac{2}{\pi \alpha}}\frac{\cosh \alpha \rho} {\rho} + {C}(z)\sqrt{\frac{\pi}{2 \alpha}}\frac{\e^{-\alpha \rho}}{\rho}+ G(\rho, s| \rho_0), \ R_1<\rho <R_2,
\end{eqnarray}
and
 \begin{eqnarray}
 \label{cQ}
    \calQ(\rho,z, s|\rho_0) =E(z)\frac{\sinh \beta \rho}{\rho}, \ 0<\rho < R_1.
\end{eqnarray}
Substituting (\ref{cP}) into the boundary condition at $\rho=R_2$ implies that
\begin{equation}
B(z)=\frac{\pi(\alpha R_2+1)}{(\alpha R_2-1)\e^{2\alpha R_2} -(\alpha R_2+1)}C(z)\equiv \frac{\pi}{\Lambda_{\alpha}(R_2)}C(z),
\end{equation}
with $\Lambda_{\alpha}(R_2)$ defined by equation (\ref{alpG1}).
The continuity conditions then give
\begin{eqnarray}
\fl & \sqrt{\frac{\pi}{2 \alpha}}{C}(z)\left [ \frac{2}{\Lambda_{\alpha}(R_2)}\frac{\cosh \alpha R_1} {R_1} + \frac{\e^{-\alpha R_1}}{R_1}\right ]
=E(z)\frac{\sinh \beta R_1}{R_1},\\
\fl &\sqrt{\frac{\pi}{2 \alpha}} {C}(z)\left [ \frac{2}{\Lambda_{\alpha}(R_2)}\left (\alpha \frac{\sinh \alpha R_1} {R_1} -\frac{\cosh \alpha R_1} {R_1^2}\right )- \frac{\e^{-\alpha R_1}}{R_1}\left (\alpha +\frac{1}{R_1}\right )\right ] +\frac{1}{4\pi DR_1^2} \widetilde{J}_{\infty}(\rho_0,s)\nonumber \\
\fl &=E(z)\left (\beta\frac{\cosh \beta R_1}{R_1}-\frac{\sinh \beta R_1}{R_1^2}\right ),
\end{eqnarray}
Again $\widetilde{J}_{\infty}(\rho_0,s)$ denotes the flux into the surface of a totally absorbing spherical target.
Rearranging these equations yields the following results
\begin{eqnarray}
E(z)&=\sqrt{\frac{\pi}{2 \alpha}}\frac{1}{\sinh \beta R_1}\left [ \frac{2}{\Lambda_{\alpha}(R_2)}\cosh \alpha R_1 + \e^{-\alpha R_1}\right ]C(z)\nonumber \\
&\equiv \sqrt{\frac{\pi}{2 \alpha}}\frac{\Theta_{\alpha}(R_1,R_2)}{\sinh \beta R_1}C(z),
\end{eqnarray}
and
\begin{eqnarray}
\fl \sqrt{\frac{\pi}{2 \alpha}}\left \{\Theta_{\alpha}(R_1,R_2) \frac{\beta R_1 \cosh \beta R_1-\sinh \beta R_1}{\sinh \beta R_1} -\Phi_{\alpha}(R_1,R_2)\right \}C(z)=\frac{1}{4\pi D}\widetilde{J}_{\infty}(\rho_0,s),\nonumber \\
\fl
\end{eqnarray}
with
\begin{equation}
\fl \Phi_{\alpha}(R_1,R_2)=\frac{2}{\Lambda_{\alpha}(R_2)}\left (\alpha R_1\sinh \alpha R_1 - \cosh \alpha R_1 \right )-\e^{-\alpha R_1} \left (\alpha R_1+1\right ).
\end{equation}
Note that the function $\Lambda_{\alpha}(R_1,R_2)$ of equation (\ref{alpG2}) can be expressed as
\begin{equation}
\Lambda_{\alpha}(R_1,R_2)=-\frac{\Phi_{\alpha}(R_1,R_2)}{R_1\Theta_{\alpha}(R_1,R_2)}.
\end{equation}
 
Combining our various results shows that within the target ($0<\rho <R_1)$,
\begin{eqnarray}
\label{qir2k}
 \calQ(\rho,z, s|\rho_0) &= E(z)\frac{\sinh \beta \rho}{\rho} =\frac{\widetilde{J}_{\infty}(\rho_0,s)}{4\pi D}\Lambda_{\alpha,\beta}(R_1,R_2)\frac{\sinh \beta \rho}{\rho},
\end{eqnarray}
with
\begin{eqnarray}
\fl \Lambda_{\alpha,\beta}(R_1,R_2)=\frac{1}{ \beta R_1 \cosh \beta R_1-\sinh \beta R_1 +R_1 \Lambda_{\alpha}(R_1,R_2)\sinh \beta R_1 }.
\end{eqnarray}
Substituting into equation (\ref{Ktil}) and introducing spherical polar coordinates shows that the Laplace-transformed flux is
\begin{eqnarray}
\fl \widetilde{K}(\rho_0,s)&=4\pi \int_0^{\infty}\psi(a) \left [\int_{0}^{R_1}\rho^2\left [\calQ(\rho,z,s|\rho_0)\right ]d\rho \right ]da\nonumber \\
\fl &=\frac{\widetilde{J}_{\infty}(\rho_0,s)}{D} \int_0^{\infty}\psi(a) {\mathcal L}_a^{-1}\left [\Lambda_{\alpha,\beta}(R_1,R_2)\int_{0}^{R_1}\rho \sinh(\beta \rho) d\rho\right ],
\label{K2}
\end{eqnarray}
assuming that we can reverse the order of integrating with respect to $\rho$ and taking the inverse Laplace transform. In the limit $s\rightarrow 0$, we have
\begin{equation}\lim_{s\rightarrow 0} \Lambda_{\alpha,\beta}(R_1,R_2)=\frac{1}{ \beta_0 R_1 \cosh \beta_0 R_1-\sinh \beta_0 R_1 },\quad \beta_0=\sqrt{\frac{z}{D}}
\end{equation}
In addition,
\begin{equation}
\int_0^{R_1}\rho \sinh(\beta \rho_0) d\rho= \frac{1}{\beta_0^2} \left (\beta_0 R_1 \cosh \beta_0 R_1-\sinh \beta_0 R_1\right ).
\end{equation}
Therefore,
\begin{eqnarray}
\lim_{s\rightarrow 0}\widetilde{K}(\rho_0,s)&=\frac{1}{D}\int_0^{\infty}\psi(a) {\mathcal L}_a^{-1}\left [\frac{D}{z} \right ]da=\int_0^{\infty} \psi(a)da=1,
\end{eqnarray}
as required.

 Finally, differentiating equation (\ref{K2}) with respect to $s$ and using equation (\ref{MFPT2}), we obtain the result
\begin{eqnarray}
\fl & T(\rho_0)=-\left .\frac{\partial}{\partial s}\widetilde{K}(\rho_0,s)\right |_{s=0}\\
\fl &=T_{\infty}(\rho_0)- \frac{1}{D}\left .\frac{\partial}{\partial s}\right |_{s=0}\int_0^{\infty}\psi(a) {\mathcal L}_a^{-1}\left [\frac{1}{\beta^2}\widehat{\Lambda}_{\alpha,\beta}(R_1,R_2) \right ]\nonumber\\
\fl &=T_{\infty}(\rho_0)-\frac{1}{D}\lim_{s\rightarrow 0} \int_0^{\infty}\psi(a) {\mathcal L}_a^{-1}\left [\frac{1}{2\sqrt{sD}}\frac{d}{d\alpha}+\frac{1}{2\sqrt{[s+z]D}}\frac{d}{d\beta}\right ]\left [\frac{1}{\beta^2}\widehat{\Lambda}_{\alpha,\beta}(R_1,R_2) \right ]da\nonumber
\end{eqnarray}
where $T_{\infty}(\rho_0)$ is the MFPT in the case of a totally absorbing target, and
\begin{equation}
\widehat{\Lambda}_{\alpha,\beta}(R_1,R_2)=\frac{1}{ 1+  R_1 \widehat{\Lambda} (\beta R_1)\Lambda_{\alpha}(R_1,R_2)  },
\end{equation}
with
\begin{equation}
\widehat{\Lambda}(r)=\frac{\sinh r}{r \cosh r-\sinh r}.
\end{equation}
Using the $\alpha$ expansion of $\Lambda_{\alpha}$, see equation (\ref{aexp}), we have
\begin{eqnarray}
\frac{d}{d\alpha}\widehat{\Lambda}_{\alpha,\beta}(R_1,R_2)&=-\frac{R_1\widehat{\Lambda}(\beta R_1)}{ \left (1+   R_1\widehat{\Lambda} (\beta R_1)\Lambda_{\alpha}(R_1,R_2) \right )^2 }\frac{d}{d\alpha} {\Lambda}_{\alpha,\beta}(R_1,R_2)\nonumber \\
 &\sim -\widehat{\Lambda}(\beta_0 R_1)\frac{2\alpha}{3R_1}(R_2^3-R_1^3)+O(\alpha^2).
\end{eqnarray}
and
\begin{eqnarray}
\fl \frac{d}{d\beta}\left (\frac{1}{\beta^2}\widehat{\Lambda}_{\alpha,\beta}(R_1,R_2)\right )&=-\frac{1}{\beta^2}\frac{R_1\Lambda_{\alpha}(R_1,R_2)}{ \left (1+ R_1  \widehat{\Lambda} (\beta R_1)\Lambda_{\alpha}(R_1,R_2) \right )^2 }\frac{d}{d\beta} \widehat{\Lambda} (\beta R_1)\nonumber \\
 \fl &\quad -\frac{2}{\beta^3}\widehat{\Lambda}_{\alpha,\beta}(R_1,R_2) \sim -\frac{2}{\beta_0^3}+O(\alpha^2).
\end{eqnarray}
Hence,
\begin{eqnarray}
\fl  T(\rho_0)&=T_{\infty}(\rho_0)+ \int_0^{\infty}\psi(a) {\mathcal L}_a^{-1}\bigg [\frac{1}{zD}\widehat{\Lambda}(\beta_0 R_1)\frac{1}{3R_1}(R_2^3-R_1^3)+\frac{1}{z^2} \bigg ] da \\
\fl &=T_{\infty}(\rho_0)+\int_0^{\infty} a \psi(a)da+\frac{1}{3DR_1}(R_2^3-R_1^3) \int_0^{\infty}\psi(a) {\mathcal L}_a^{-1}\frac{1}{z}\widehat{\Lambda}(R_1\sqrt{z/D})da.\nonumber
\end{eqnarray}

One immediate result is that in the limit $R_2\rightarrow R_1$, the particle spends all of its time within the interior of the target and
\begin{equation}
T(\rho_0)\rightarrow \int_0^{\infty}a\psi(a)da\equiv \E[a].
\end{equation}
We have used the fact that $T_{\infty}(\rho_0)\rightarrow 0$ as $R_2\rightarrow R_1$ since $\rho_0\rightarrow R_1$, which means that the particle starts on the totally absorbing boundary. This is one major difference from the case of Sect. 4.1, where $T(\rho_0)\rightarrow 0$ as $R_2\rightarrow R_1$.
Now suppose that $R_2>R_1$ and $\psi(a)$ has finite moments with $R_1\gg \sqrt{D\E[a]}\gg \sqrt{D\E[a^n]} $ for all integers $n>1$. Then
\begin{eqnarray}
\fl \int_0^{\infty}\psi(a) {\mathcal L}_a^{-1}\frac{1}{z}\widehat{\Lambda}(R_1\sqrt{z/D})da=\frac{R_1^2}{D}\int_0^{\infty}\psi(a){\mathcal L}_a^{-1}  \widetilde{f}(R_1^2z/D)da,
\end{eqnarray}
where $f(r)=\widehat{\Lambda}(r)/r$. Moreover, 
\[ \widetilde{f}(R_1^2z/D)=\int_0^{\infty} \e^{-R_1^2 za'/D}f(a') da'=\frac{D}{R_1^2} \int_0^{\infty} \e^{-za}f(aD/R_1^2) da.\]
Hence,
\begin{eqnarray}
\fl \int_0^{\infty}\psi(a){\mathcal L}_a^{-1}  \widetilde{f}(R_1^2z/D)da=\frac{D}{R_1^2}\int_0^{\infty}\psi(a) f(aD/R_1^2) da
\end{eqnarray}
Since, $\psi(a)$ is dominated in regions where $aD/R_1^2 \ll1$, it follows that the integral on the right-hand side depends on the behavior of $f(r)$ near $r=0$, which itself is determined by the large-$z$ behavior of $\widetilde{f}(z)$. Noting that $\widetilde{f}(z)\sim 3/z^2$ as $z\rightarrow \infty$, we deduce that
\begin{eqnarray}
\fl \int_0^{\infty}\psi(a) {\mathcal L}_a^{-1}\frac{1}{z}\widehat{\Lambda}(R_1\sqrt{z/D})da\approx
\frac{3D}{R_1^2}\int_0^{\infty}\psi(a) {\mathcal L}_a^{-1}\left [\frac{1}{z^2}\right ]da = \frac{3D}{R_1^2}\int_0^{\infty}\psi(a) ada.
\end{eqnarray}
Hence,
\begin{equation}
\fl T(\rho_0)\approx T_{\infty}(\rho_0)+\E[a]+\frac{1}{R_1^3}(R_2^3-R_1^3) \E[a]=T_{\infty}(\rho_0)+\left (\frac{R_2}{R_1}\right )^3 \E[a].
\end{equation}
Hence, as in the analysis of the stopping local time, we see that the MFPT blows up if the first moment of the stopping occupation time is infinite. Finally, comparison with equation (\ref{Tfin3D}) shows that for finite first moments
\begin{eqnarray}
\frac{[T(\rho_0)-T_{\infty}(\rho_0)]_{\rm boundary}}{[T(\rho_0)-T_{\infty}(\rho_0)]_{\rm interior}}
\approx\frac{ R_1[R_2^3-R_1^3]}{3DR_2^3}\frac{\E[\ell]}{\E[a]}.
\end{eqnarray}
In the particular case of constant reaction rates, $\E[\ell]=D/\kappa_0$ and $\E[a]=1/k_0$. (Recall that $\ell$ has units of length, whereas $a$ has units of time.) Fixing these rates according to $R_1k_0=\kappa_0$, shows that the ratio of $T(\rho_0)-T_{\infty}(\rho_0)$ for boundary and interior absorption is given by the geometrical factor
\begin{equation}
\chi(R_1,R_2)=\frac{1}{3}\left (1-\left [\frac{R_1}{R_2}\right ]^3\right ),\quad R_1 \ll \sqrt{D/k_0}.
\end{equation}

 \setcounter{equation}{0}
\section{Propagator BVP for multiple targets} 

Another advantage of the Feynman-Kac approach is that it is relatively straightforward to extend the theory to multiple targets, each with its own local surface reaction scheme. In order to show this, we will focus on the case of absorption at the target boundaries. However, a very similar construction can be carried out in the case of absorption within the target interiors. Suppose that the domain $\Omega$ contains $N$ targets $\calU_j$, $j=1,\ldots,N$, with partially reactive surfaces $\partial \calU_j$. Let $\ell_{j,t}$ denote the local time of the $j$th target with
\begin{equation}
\label{mloc}
\ell_{j,t}=\lim_{h\rightarrow 0} \frac{D}{h} \int_0^t\Theta(h-\mbox{dist}(\X_{\tau},\partial \calU_j))d\tau,
\end{equation}
and $\X_t$ representing the position of a particle undergoing reflected Brownian motion in $\Omega\backslash \cup_{j=1}^N \calU_j$.
Consider the generalized propagator $P=P(\x,{\bm \ell},t|\x_0)$, ${\bm \ell}=(\ell_1,\ldots\ell_N)$, for the set of random variables $(\X_t,\ell_{1,t},\ldots,\ell_{N,t})$. For each target introduce the stopping time
\begin{equation}
\label{mTell}
\tau_j=\inf\{t>0:\ \ell_{j,t} >\widehat{\ell}_j\},\quad j=1,\ldots,N,
\end{equation}
where $\widehat{\ell}_j$ is the corresponding stopping local time with distribution $\Psi_j(\ell_j)$. We then define the FPT to be
\begin{equation}
{\mathcal T}=\min\{\tau_1,\tau_2,\ldots,\tau_N\}.
\end{equation}
Since the stopping local times $\widehat{\ell}_j$ are statistically independent, the relationship between $p(\x,t|\x_0)$ and $P(\x,\ell_1,\ldots,\ell_N,,t|\x_0)$ can be established as follows:
\begin{eqnarray*}
\fl p(\x,t|\x_0)d\x&=\P[\X_t \in (\x,\x+d\x), \ t < {\mathcal T}|\X_0=\x_0]\\
\fl &=\P[\X_t \in (\x,\x+d\x), \ {\bm \ell}_{t }< \widehat{\bm \ell}|\X_0=\x_0]\\
\fl &=\int_0^{\infty} d\ell_1 \psi_1(\ell_1)\cdots\int_0^{\infty} d\ell_N \psi_1(\ell_N) \P[\X_t \in (\x,\x+d\x), \ {\bm \ell}_t < {\bm \ell} |\X_0=\x_0]\\
\fl &=\int_0^{\infty} d\ell_1 \psi_1(\ell_1)\cdots\int_0^{\infty} d\ell_N \psi_1(\ell_N) \int_0^{\ell_1} d\ell'_1 \cdots \int_0^{\ell_N}d\ell_N'[P(\x,{\bm \ell}',t|\x_0)d\x].
\end{eqnarray*}
Reversing the orders of integration and setting $\psi_j=-\Psi_j'$ yields the result
\begin{equation}
\label{mbob}
p(\x,t|\x_0)=\int_0^{\infty}d\ell_1\Psi(\ell_1)\cdots \int_0^{\infty} d\ell_N\Psi_N(\ell_N) P(\x,{\bm \ell},t|\x_0).
\end{equation}

We can now derive a BVP for the propagator by noting that
 \begin{eqnarray}
 \label{mA1}
& P(\x,{\bm \ell},t|\x_0)=\bigg \langle \prod_{j=1}^N\delta\left (\ell_j -\ell_{j,t }\right )\bigg \rangle_{\X_0=\x_0}^{\X_t=\x} ,
 \end{eqnarray}
 where expectation is again taken with respect to all random paths realized by $\X_{\tau}$ between $\X_0=\x_0$ and $\X_t=\x$. 
  Using a Fourier representation of each Dirac delta function, equation (\ref{mA1}) can be rewritten as
 \begin{eqnarray}
 \fl P(\x,{\bm \ell},t|\x_0)=\int_{-\infty}^{\infty}\frac{d\omega_1}{2\pi}\cdots \int_{-\infty}^{\infty}\frac{d\omega_N}{2\pi} \e^{i\sum_{j=1}^N\omega_j \ell_j}{\mathcal G}(\x,{\bm \omega},t|\x_0),
 \end{eqnarray}
 where ${\bm \omega}=(\omega_1,\ldots,\omega_N)$, $ P(\x,{\bm \ell},t|\x_0)=0$ if $\ell_k <0$ for at least one value of $k$, and
 \begin{eqnarray}
 {\mathcal G}(\x,{\bm \omega},t|\x_0)=\bigg\langle \exp \left ( -i\sum_{j=1}^N\omega_j \ell_{j,t}\right )\bigg \rangle_{\X_0=\x_0}^{\X_t=\x}.
 \end{eqnarray}
 The corresponding Feynman-Kac equation is
\begin{eqnarray}
\label{mcalG}
\fl \frac{\partial \calG(\x,{\bm \omega},t|\x_0)}{\partial t}&=D\nabla^2 \calG(\x,{\bm \omega},t|\x_0) -i\sum_{j=1}^N\omega_j  F_j(\x) \calG(\x,{\bm \omega},t|\x_0) 
\end{eqnarray}
for $\x \in \Omega\backslash \cup_{j=1}^N \calU_j$, with
\begin{eqnarray}
F_j(\x)= D\int_{\partial \calU_j}\delta(\x-\x')d\x'.
\end{eqnarray}
Multiplying equation (\ref{mcalG}) by $\e^{\i{\bm \omega}\cdot {\bm  \ell}}$ and integrating with respect to ${\bm \omega}$ gives
\begin{eqnarray}
\fl \frac{\partial P(\x,{\bm \ell},t|\x_0)}{\partial t}&=D\nabla^2 P(\x,{\bm \ell},t|\x_0)\\
\fl &\quad -D\sum_{j=1}^N\int_{\partial \calU_j}\left (\frac{\partial P}{\partial \ell_j}(\x',\ell,t|\x_0) +\delta(\ell_j)P(\x',{\bm \ell},t|\x_0) \right )\delta(\x-\x')d\x' ,\nonumber 
\end{eqnarray}
together with a no-flux boundary condition on $\partial \Omega$.
This is equivalent to the BVP
\numparts
\begin{eqnarray}
\label{mPloc1}
\fl &\frac{\partial P(\x,{\bm \ell},t|\x_0)}{\partial t}=D\nabla^2 P(\x,{\bm \ell},t|\x_0) \mbox{ for }  \x \in \Omega\backslash \cup_{j=1}^N \calU_j, \\ \fl  &\nabla P(\x,{\bm \ell},t|\x_0) \cdot \n =0 \mbox{ for } \x\in \partial \Omega,\\
\fl &-D\nabla P(\x,{\bm \ell},t|\x_0) \cdot \n_k= D P(\x,{\bm \ell},t|\x_0) \ \delta(\ell_k)  +D\frac{\partial}{\partial \ell_k} P(\x,{\bm \ell},t|\x_0) 
\label{mPloc2}
\end{eqnarray}
for $\x\in \partial \calU_k$, $k=1,\ldots,N$, with
\begin{equation}
\label{mPloc3}
P(\x,{\bm \ell},t|\x_0)|_{\ell_k=0}=-\nabla p_{k,\infty}(\x,t|\x_0)\cdot \n_k \mbox{ for }\x\in \partial \calU_k, 
\end{equation}
\endnumparts
where $p_{k,\infty}$ is the probability density in the case of a single totally absorbing target $\calU_k$ in the bounded domain $\Omega_k=\Omega\backslash \cup_{j\neq k}\calU_j$:
\numparts 
\begin{eqnarray}
\label{mpinf}
\fl 	&\frac{\partial p_{k,\infty}(\x,t|\x_0)}{\partial t} = D\nabla^2 p_{k,\infty}(\x,t|\x_0) \mbox{ for } \, \x\in \Omega\backslash \cup_{j\neq k}\calU_j,\\ \fl &  \nabla p_{k,\infty}(\x,t|\x_0) \cdot \n=0 \mbox{ for } \x\in\partial \Omega,\\
\fl &p_{k,\infty}(\x,t|\x_0)=0 \mbox{ for } \x\in \partial \calU_k,\  \nabla p_{j,\infty}(\x,t|\x_0) \cdot \n_j=0 \mbox{ for } \x\in\partial \calU_j,\ j\neq k .
	\end{eqnarray}
	\endnumparts 
	Here $\n_k$ denotes the inward unit normal of the $k$th target.
	
Once the propagator has been determined, the marginal probability density $p(\x,t|\x_0)$ for particle position can be obtained using equation (\ref{mbob}). The associated flux into the $k$th target is 
\begin{eqnarray}
J_k(\x_0,t)=-D \int_{ \partial \calU_k}\nabla p(\x,t|\x_0)\cdot \n_k d\sigma .
\label{mflux}
\end{eqnarray}
The survival probability that the particle hasn't been absorbed by any of the targets in the time interval $[0,t]$, having started at $\x_0$, is defined according to
\begin{equation}
\label{mS1}
S(\x_0,t)=\int_{\Omega\backslash \cup_{j=1}^N\calU_j}p(\x,t|\x_0)d\x.
\end{equation}
Differentiating both sides of this equation with respect to $t$ and using the diffusion equation
gives
\begin{eqnarray}
\fl \frac{\partial S(\x_0,t)}{\partial t}&=D\int_{\Omega\backslash \cup_{j=1}^N\calU_j} \nabla\cdot \nabla p(\x,t|\x_0)d\x=D \sum_{j=1}^N\int_{ \partial \calU_j}\nabla p(\x,t|\x_0)\cdot \n_j d\sigma\nonumber \\
\fl & =-\sum_{j=1}^NJ_j(\x_0,t),
\label{mQ2}
\end{eqnarray}
Let ${\mathcal T}_k(\x_0)$ denote the FPT that the particle is captured by the $k$-th target, with ${\mathcal T}_k(\x_0)=\infty$ indicating that it is not captured by that specific target.
Let $\Pi_k(\x_0,t)$ denote the probability that the particle is captured by the $k$-th target after time $t$, given that it started at $\x_0$:
\begin{equation}
\Pi_k(\x_0,t)=\P[t<{\mathcal T}_k(\x_0)<\infty ]=\int_t^{\infty} J_k(\x_0,t')dt',
\end{equation}
The corresponding FPT density is $f_k(\x_0,t)=J_k(\x_0,t)$.
The splitting probability $\pi_k(\x_0)$ and conditional MFPT $T_k(\x_0)$ to be captured by the $k$-th target are then defined according to
\begin{equation}
\pi_k(\x_0)  \equiv \Pi_k(\x_0,0)= \int_0^\infty J_k(\x_0,t) dt=\widetilde{J}_k(\x_0,0),
\end{equation}
and
\begin{eqnarray}
\fl T_k(\x_0) &\equiv  \mathbb{E}[{\mathcal T}_k | {\mathcal T}_k < \infty]=\frac{1}{\pi_k(\x_0)}\int_0^{\infty} t J_k(\x_0,t) dt =-\frac{1}{\pi_k(\x_0)}\left .\frac{\partial}{\partial s}\widetilde{J}_k(\x_0,s)\right |_{s=0}.
\end{eqnarray}

Finally, the Laplace transformed fluxes can be expressed directly in terms of the propagator using the boundary condition (\ref{mPloc2}). Multiplying both sides of the latter by $\prod_{i=1}^N\Psi_i(\ell_i)$ and integrating by parts with respect to $\ell$ shows that
\begin{eqnarray}
\fl -D\nabla p(\x,t|\x_0) \cdot \n_k=D\int_0^{\infty}\psi_k(\ell_k)\left [ \prod_{j\neq k} d\ell_j \Psi_j(\ell_j)\right ]P(\x,{\bm \ell},t|\x_0)d\ell 
\end{eqnarray}
for $\x\in \partial \calU_k,$.
Integrating with respect to points on the boundary $\partial \calU_k$ and Laplace transforming gives
\begin{equation}
\label{mJ}
\fl \widetilde{J}(\x_0,s)=D\int_0^{\infty}d\ell_k \psi_k(\ell_k)\left [ \prod_{j\neq k} d\ell_j \Psi_j(\ell_j)\right ] \int_{\partial \calU_k}\PP(\x,{\bm \ell},s|\x_0)d\sigma.
\end{equation}

\section{Conclusion}  

In this paper we developed a unified probabilistic framework for analyzing diffusion-mediated surface reactions, which applies irrespective of whether absorption occurs at the boundary or within the interior of a chemically active target substrate. We proceeded by using the Feynman-Kac formula to derive a BVP for the joint probability density (propagator) of particle position and a general Brownian functional. Absorption at the boundary or interior of a single target was then modeled by taking the Brownian functional to be the boundary local time or the occupation time, respectively, and introducing a corresponding stopping condition. We applied the theory to the case of a concentric spherical shell whose interior surface was partially reactive and whose outer surface was totally reflecting. In particular, we calculated the MFPT for absorption and showed that the MFPT diverged if the probability density of the stopping local or occupation time had an infinite first moment. We also illustrated how to calculate the propagator directly by solving its BVP, rather than using the spectral decomposition of an associated Dirichlet-Neumann operator \cite{Grebenkov20}. Finally, we further extended the theory to the case of multiple, non-identical targets by introducing a separate local or occupation time for each target. The associated propagator BVP was also derived using the Feynman-Kac formula. The analytical framework developed in this paper could be used to investigate the competition for resources between multiple partially reactive targets. A specific application in cell biology  would be the transport and delivery of proteins to neuronal synapses, whose interiors act as reactive surfaces. Since it is non-trivial to obtain an exact solution of the propagator BVP in the case of multiple targets, some form of approximation scheme would be needed. For example, in the small-target limit one could adapt asymptotic methods previously used to solve the so-called narrow capture problem  for totally absorbing targets \cite{Coombs09,Chevalier11,Ward15,Lindsay16,Bressloff21a,Bressloff21b}.

\end{document}